\documentclass[aps,twocolumn]{revtex4} 
\input epsf

\newcommand\lsim{\mathrel{\rlap{\lower4pt\hbox{\hskip1pt$\sim$}}
        \raise1pt\hbox{$<$}}}
\newcommand\gsim{\mathrel{\rlap{\lower4pt\hbox{\hskip1pt$\sim$}}
        \raise1pt\hbox{$>$}}}  

\newcommand{\sfig}[2]{
\centerline{ \epsfxsize = #2 \epsfbox{#1} }
		}
\newcommand{\Sfig}[2]{
	\begin{figure}[t]
	\sfig{#1.eps}{0.95\columnwidth}
	\caption{{\small #2}}
	\label{fig:#1}
	\end{figure}
}

\def\cmm2{{\,\rm cm^{-2}}}
\def\cm2{{\,{\rm cm}^2}}
\def\cmm3{{\,{\rm cm}^{-3}}}
\def\gcmm3{{\,{\rm g\,cm^{-3}}}}

\def\fun#1#2{\lower3.6pt\vbox{\baselineskip0pt\lineskip.9pt
  \ialign{$\mathsurround=0pt#1\hfil##\hfil$\crcr#2\crcr\sim\crcr}}}

\def\be{\begin{equation}}
\def\ee{\end{equation}}
\def\bea{\begin{eqnarray}}
\def\eea{\end{eqnarray}}

\newcommand{\ec}[1]{Eq.~(\ref{eq:#1})}

\begin{document}

\title{A Horizon Ratio Bound for Inflationary Fluctuations}

\author{Scott Dodelson and Lam Hui
}

\affiliation{NASA/Fermilab Astrophysics Center,
Fermi National Accelerator Laboratory, Batavia, IL~~60510}
\affiliation{Department of Astronomy \& Astrophysics, University of Chicago, 
Chicago, IL 60637 
}

\date{\today}

\begin{abstract}
We demonstrate that the gravity wave background amplitude
implies a robust upper bound on the ratio: $\lambda / H^{-1} \lsim e^{60}$, 
where $\lambda$ is the proper wavelength of fluctuations of interest and 
$H^{-1}$ is the horizon at the end of inflation. The bound holds
as long as the energy density of the universe does not drop faster than
radiation subsequent to inflation. 
This limit implies that the amount of expansion between the time the scales of
interest leave the horizon and the end of inflation, denoted by $e^{N}$,
is also bounded from above, by about $e^{60}$ times a factor that involves
an integral over the first slow-roll parameter. In other words, the bound
on $N$ is model dependent -- we show that for vast classes of slow-roll models, 
$N \lsim 67$. The quantities, $\lambda / H^{-1}$ or $N$, play an important
role in determining the nature of inflationary scalar and tensor fluctuations. We suggest 
ways to incorporate the above bounds when confronting inflation models with observations. 
As an example, this bound solidifies the tension between observations 
of cosmic microwave background (CMB) anisotropies and 
chaotic inflation with a $\phi^4$ potential by closing the escape hatch of
large $N$ ($< 62$).
\vskip 0.3truecm
98.80.Bp; 98.80.Cq; 98.65.Dx
\end{abstract}
\maketitle

The purpose of this short note is two-fold. First, 
we briefly review how fluctuations
predicted by inflation \cite{oldandnewinflation} are related to $N$, the  
number of e-folds between the time
the scales of interest leave the horizon 
\cite{horizon} and the end of inflation (\S \ref{review}). (Note that
$N$ is {\it not} the total number of e-folds of inflation, a
generally bigger number.) Second, we derive
an upper bound on $N$ which should be used when constraining inflationary models. 
This is done in two steps:
\begin{itemize}
\item we derive a robust, model-independent 
limit on $\tilde N$, defined to be the logarithm of the ratio of the
proper wavelength of cosmological modes to the 
horizon at the end of inflation (\S \ref{secNtilde}).
\item From this, we infer
a model-dependent bound on $N$ (\S \ref{secN}).
\end{itemize}
It is not uncommon to find in the literature a wide
variety of assumptions made about $N$, and we find it timely to point out the
importance of this bound, especially in light of improving observations.
It should be emphasized that while a fair fraction of our discussion
is confined to single-field slow-roll inflation for the sake of simplicity,
the constraint on $\tilde N$ in \S \ref{secNtilde} is quite 
general, applicable to a much wider variety of inflation models.
This leads to a short discussion in \S \ref{discuss} where we observe that
$\tilde N$ might be a better independent variable to adopt instead of $N$,
when solving the inflationary flow equations.

While revision of this paper was under way, a paper
by Liddle and Leach \cite{ll} appeared which reached very 
similar conclusions.

\section{A Brief Review}
\label{review}

For large classes of single-field, slow-roll inflationary models, the 
predictions for
scalar and tensor fluctuations can be summarized as follows (to lowest order in
slow-roll) \cite{manypapers}:
\begin{eqnarray}
\label{basics}
n_s - 1 = \sigma \quad , \quad r = - n_T/2 = \epsilon
\end{eqnarray}
where $n_s$ is the scalar spectral index, $r$ is the tensor to scalar ratio, and
$n_T$ is the tensor spectral index. The equality $r = -n_T/2$ expresses
the well-known consistency relation \cite{consistency}.

The slow-roll parameters $\epsilon$ and $\sigma$ are related to derivatives
of the Hubble parameter $H$ as a function of inflaton field value $\phi$:
\begin{eqnarray}
\label{epsilon}
\epsilon \equiv {m_{\rm pl.}^2 \over 4\pi} \left({H' \over H}\right)^2 
\quad , \quad 
\sigma \equiv {m_{\rm pl.}^2 \over 2 \pi} \left[ {H'' \over H}
- 2 \left({H' \over H}\right)^2 \right]
\end{eqnarray}
where the prime denotes differentiation with respect to $\phi$. 
For a wave-mode of interest, equation (\ref{basics})
is to be evaluated at horizon crossing during inflation. This is equivalent
to evaluating equation (\ref{epsilon}) at the corresponding field value
$\phi=\phi_*$ (hereafter $*$ is used to denote the time of horizon exit), or,
as is commonly done, at the corresponding $N$:
\begin{eqnarray}
\label{Nfold}
N(\phi_*) \equiv \int_{t_*}^{t_e} dt H = 
{\sqrt{4\pi} \over m_{\rm pl.}} \, \left| \int_{\phi_*}^{\phi_e} d\phi / \sqrt{\epsilon} 
\right|
\end{eqnarray}
where $t$ is the proper time. Here $N$ is the number of e-folds between the horizon exit of the scale of interest 
(i.e. $t_*$ or $\phi_*$) and the end of inflation ($t_e$ or $\phi_e$). 
The end of inflation is defined to be the time when slow-roll ends. 

A hierarchy of flow equations tells us how the slow-roll parameters depend on $N$
\cite{flow}:
\begin{eqnarray}
\label{flow}
&& {d\epsilon \over dN} = \epsilon (\sigma + 2\epsilon) \, , \,
{d\sigma \over dN} = -5 \epsilon \sigma - 12 \epsilon^2 + 2 (^2\lambda) \, ,\\ \nonumber 
&& {d(^\ell \lambda)  \over dN} = [(\ell - 1)\sigma/2 + (\ell - 2)\epsilon] (^\ell \lambda) + \,^{\ell+1}\lambda
\end{eqnarray}
where $\ell$ ranges from $2$ to in principle infinity, and
$^\ell \lambda$'s are the higher order slow-roll parameters.
In understanding the dynamics of inflation, it is also useful to remember
the equation of motion for $\phi$: 
$\dot \phi = - m_{\rm pl.}^2 H' /(4\pi)$, and 
the Friedmann equation: $3 H^2 = (8\pi/m_{\rm pl.}^2)
[V + \dot\phi^2/2]$, where $\dot\phi$ is the derivative of $\phi$
with respect to proper time, and $V$ is the inflaton potential \cite{mc}.

As expressed above, it is clear that $N$ plays an important role in determining
the properties of observable fluctuations. One can imagine a bound on $N$ provides
useful information about the fluctuations, although the precise manner
depends on the particular model under consideration. To take a simple example,
for chaotic inflation with a $\phi^p$ potential:
$N(\phi_*) + p/4 = 4 \pi \phi_*^2 /(p m_{\rm pl.}^2)$ (where we have used the fact
that $\epsilon = 1$ at the end of inflation), 
and $\epsilon = p/(p+4N)$, $\sigma = -(2+p)/(2N+p/2)$, leading to
(at the lowest order):
\begin{eqnarray}
\label{chaoticN}
n_s - 1 = -(2+p)/(2N+p/2) \quad , \quad r =  p/(p+4N)
\end{eqnarray}

The predictions of chaotic inflation then are quite sensitive to the precise
value of $N$, and this dependence holds for many inflationary models
\cite{othermodels}.
This leads to an important question: what are the constraints on 
$N$? The WMAP team \cite{wmap} fixed $N$ to be $50$ and then proceeded to
show that their data excluded the $\phi^4$ model. Reference~\cite{barger}
pointed out though that $N$ need not be fixed at $50$, and loosening this
constraint correspondingly loosens the constraints on the $\phi^4$ chaotic
inflation model. It is not uncommon in the literature 
to allow $N$ to range up to $70$ (e.g. \cite{whk}).

\section{A Model-independent Bound on $\tilde N$}
\label{secNtilde}

First, we derive a bound on a slightly different quantity, which
turns out to be more robust. Let us \cite{lpb} define
$e^{\tilde N} \equiv a_e H_e/k$, where $k$ is the comoving
wavenumber of interest, $a_e$ is the scale factor
and $H_e$ is the Hubble parameter, both at the end of inflation.
Hereafter the subscript $e$ refers to the end of inflation.
In other words, $e^{\tilde N}$ is the ratio of the physical wavelength
$(a_e/k)$ to the Hubble radius $(H_e^{-1})$ at the end of inflation. 
It can be calculated backwards from today: there is a symmetry in the evolution of
$aH/k$. During inflation this ratio increases from unity at horizon crossing
to $e^{\tilde N}$, and then after inflation it falls back to unity once the
scale re-enters the horizon. 
The bound can be derived by extrapolating
backwards from today to get $a_e$ as a function of $H_e$, so that $\tilde N$ is solely
a function of $H_e$ and then arguing that $H_e$ is less than or equal to $H_*$. 

Let us now develop the argument in more detail to make sure we
arrive at a conservative bound. Naively,
one expects $H_e = H_0 \Omega_{r,0}^{1/2} a_e^{-2}$, where 
$\Omega_{r,0} = 4.2 \times 10^{-5} h^{-2}$ is the radiation density today in units of the critical
density, with $h \equiv H_0/$ ($100$ km/s/Mpc) parametrizing the Hubble
constant today. 
Taking into account changes in the number of relativistic
species, as well as the possibility of decoupled degrees of freedom (e.g.
neutrinos today), 
one should use instead $H_e = H_0 \Omega_{r,0}^{1/2} a_e^{-2}
\left[ (g_e / g_0) (g_0^S / g_e^S)^{4/3}\right]^{1/2}$. Here, $g$ is the effective degrees of freedom
that relates the energy density $\rho$ to temperature $T$: $\rho \propto g T^4$, while 
$g^S$ relates the entropy density $s$ to $T$: $s \propto g^S T^3$. If $g$ and $g^S$
were identical, then the factor in square brackets would be $(g_0/g_e)^{1/6}$, smaller
than $(3.36/100)^{1/6}=0.57$ since the standard model alone contains more than $100$ relativistic
degrees of freedom at very high temperatures. The difference between the $g$'s \cite{kolbturner}
mitigates this to some extent and is somewhat model dependent; a conservative bound 
follows from setting the coefficient to unity,
so $a_e < (H_0/H_e)^{1/2} \Omega_{r,0}^{1/4}$. Thus,
\begin{equation}
e^{\tilde N} = {a_eH_e\over k} < 0.08 \left({H_e\over H_0}\right)^{1/2} \left({H_0\over k}\right) h^{-1/2}
.
\label{eq:first}
\end{equation}
Using now the weak assumption that $H_*$, the Hubble parameter in the early part of inflation
when the fluctuation leaves the horizon, is larger than $H_e$, we arrive at
\begin{equation}
e^{\tilde N} < e^{60.9}  \left({H_*\over 10^{15} {\rm GeV}}\right)^{1/2} \left({0.002\, {\rm Mpc}^{-1} \over k}\right) .
\label{eq:nlimit}
\end{equation}
Note that $\tilde N$ is a function of scale $k$.
The scale $k = 0.002$ Mpc$^{-1}$ is well-measured by the CMB, so it is
a convenient pivot spot~\cite{wmap}.

There is one possible loophole in \ec{first}.
The end of slow-roll ($a_e$) is generally earlier than the time when
the universe finally completes reheating to become radiation dominated.
Equation (\ref{eq:first}) assumes that this transition is instantaneous,
but relaxing this assumption only strengthens the inequality.
To see this, for a given $H_e$, define a quantity $a_e^{\rm eff.}$, which
is the scale factor if one were to extrapolate backward
from the end of reheating to a time when the Hubble parameter is $H_e$, 
as if the universe remains radiation dominated between these two times.
With the weak assumption that the true Hubble parameter should fall slower
than $a^{-2}$ between these two times, one can see that 
$a_e < a_e^{\rm eff.}$. Combining this with the 
relation $a_e^{\rm eff.} < (H_0/H_e)^{1/2} \Omega_{r,0}^{1/4}$ gives
us back the inequality in \ec{first}.

The gravity wave amplitude is proportional to $H_*$.
A conservative bound ($3 \sigma$) from observations of the CMB anisotropies is
$H_* < 3.3 \times 10^{14}$ GeV \cite{tensor}. Hence, \ec{nlimit}
constrains~\cite{andrew}
\begin{equation}
\tilde N < 60 + {\,\rm ln \,} \left({0.002\, {\rm Mpc}^{-1} \over k}\right)
\label{Ncon}
.\end{equation}
The largest observable scale today corresponds to 
$k = H_0$, implying the largest possible observationally
relevant $\tilde N$ is $62 + {\,\rm ln}(0.7/h)$.

We refer to this limit on $\tilde N$ as the {\it horizon ratio bound},
as it derives from comparing 
the horizon today with that at the end of inflation.
An important assumption is that the Hubble parameter does not fall faster than
$a^{-2}$ after the end of inflation i.e. the energy density does not redshift
faster than radiation. If, for instance, there is an
extended period of domination by a kinetic-energy-dominated scalar field
($H \propto a^{-3}$), the above bound would be violated. On the other hand, periods
of late entropy production or secondary inflation would only serve
to strengthen our bound. This caveat aside, our bound is quite general --
it is independent of the exact model of inflation. 

\section{Upper Bound(s) on $N$}
\label{secN}

The amount of expansion between horizon exit and the end of inflation
is given by $e^N = a_e H_* / k = e^{\tilde N} H_*/H_e$. 
Following equation (\ref{eq:first}), we see that
\begin{eqnarray}
e^N < 0.08 \left({H_0\over k}\right) h^{-1/2}\Big[ \left({H_*\over H_e}\right)^{1/2} 
\left({H_*\over H_0}\right)^{1/2} \Big]
\end{eqnarray}
The second term inside the square brackets can be
bounded using the gravity wave amplitude as before. The first
is the square root of the ratio of the Hubble parameter at
exit and at the end of inflation. This ratio can be rewritten
using equations (\ref{epsilon}) and (\ref{Nfold}): 
$H_*/H_e$ as a function of $N$ is given by ${\rm exp\,} [\int_0^N \epsilon(N') dN']$
\cite{liddle93}.
Hence, we obtain
\begin{eqnarray}
\label{Nepsilon}
N < 60 + {1\over 2} \int_0^N \epsilon(N') dN' +  {\,\rm ln \,} \left({0.002\, {\rm Mpc}^{-1} \over k}\right)
\end{eqnarray}
The integral over $\epsilon$ introduces a dependence on the inflation model
to the bound on $N$.
The weakest statement one could make is that $\epsilon < 1$ during inflation,
and so the integral has to be less than $N$, implying
a bound on $N$ that is weaker than the one on $\tilde N$ by a factor of 2. 
Imposing the requirement that inflation has to end before nucleosynthesis
(temperature $\sim 1$ MeV) strengthens this bound somewhat to 
$N < 105 + {\,\rm ln \,} (0.002\, {\rm Mpc}^{-1} / k)$. This is 
our most general model-independent bound on $N$.

\Sfig{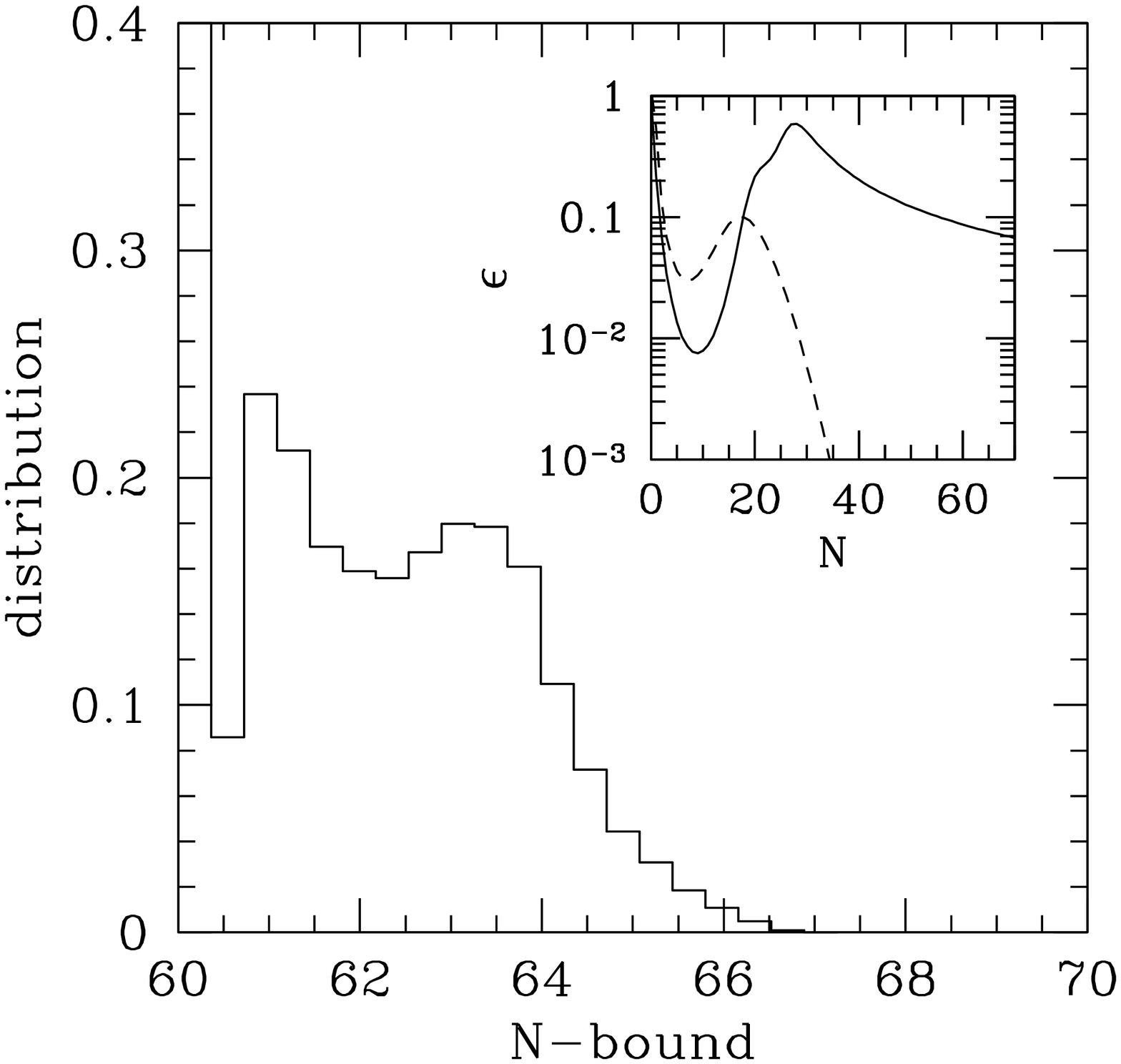}{The probability distribution of $N$-bound (eq. \ref{Nepsilon})
among a host of Monte Carlo realizations of inflation models. 
The spike around $60$ is largely due to fixed points, models where
inflation does not terminate at $\epsilon = 1$, but rather $\epsilon \sim 0$.
The inset shows two examples of how $\epsilon$ flows with $N$ (i.e. not fixed points).
}

However, generic single-field slow-roll models (including hybrid models
as effective single-field models) likely obey a significantly 
stronger bound on $N$. We perform an integration of the flow equations (equation \ref{flow})
up to the 5th order in slow-roll (i.e. $\ell = 5$), for a million
randomly generated models in the slow-roll parameter space, following the prescription of
\cite{whk}. The trajectories of $\epsilon$ can be used
to evaluate the integral in equation (\ref{Nepsilon}). We solve for the
resulting bound on $N$ for each model, 
whose probability distribution is shown in Fig.~\ref{fig:pdf}.
It appears there is an upper limit on $N$:
\begin{eqnarray}
N < 67 + {\,\rm ln \,} (0.002\, {\rm Mpc}^{-1} / k)
\end{eqnarray}
We do find, however, some evidence for a weak increase in this upper bound
as one truncates the slow-roll flow equations at higher orders. 
We therefore recommend using equation (\ref{Nepsilon}) to evaluate the
appropriate bound on a case by case basis.

An instructive example to see why the model-dependent correction to the
$N$-bound is small is chaotic inflation with a $\phi^4$ potential. 
From \S \ref{review}, we know $\epsilon = 1/(1+N)$, and so 
$\int_0^N \epsilon(N') dN' = {\,\rm ln \,} (1+N)$. 
Plugging this into equation (\ref{Nepsilon}) implies a bound of
$N < 62 + {\,\rm ln\,} (0.002 {\,\rm Mpc^{-1}\,} / k)$. 
Such a modest $N$ for the $\phi^4$ model runs the danger of 
producing too much spectral tilt and/or too high a tensor to scalar
ratio (equation \ref{chaoticN}).
Recently, \cite{kkmr} showed that the combination of
WMAP with seven other CMB experiments rules out the $\phi^4$ model 
at $3\sigma$ unless $N$ is larger than
$66$. This, together with our bound, appears to rule out $\phi^4$
chaotic inflation. However, we caution that 
\cite{kkmr} combined different experiments assuming independence.

\section{Discussion}
\label{discuss}

In summary, we have derived a model-independent upper limit of about 
$e^{\tilde N} < e^{60}$ on the 
ratio of wavelength to horizon size at the end of inflation (equation \ref{Ncon}).
A corresponding model-dependent upper limit on $e^{N}$, which is
the amount of expansion between horizon exit and the end of inflation, 
is given in equation (\ref{Nepsilon}). For vast classes of slow-roll models,
we find that this gives a bound of $N < 67$. 

The discussion so far points to two different ways of
implementing the horizon-ratio bound. One is to use equation (\ref{Nepsilon})
and evaluate the model-dependent correction on a case by case basis.
The other is to bypass the use of $N$ altogether. It can be shown from
equations (\ref{epsilon}) and (\ref{Nfold}) that
\begin{eqnarray}
(1+\epsilon) {d \over d \tilde N} = {d \over dN}
\end{eqnarray}
This can be used to rewrite the flow equations (\ref{flow}) using
$\tilde N$ instead of $N$ as the independent variable. 
The predictions for inflationary fluctuations can therefore
be expressed in terms of $\tilde N$ in place of $N$. 
Our robust bound on $\tilde N$ can be implemented
directly. We will explore this further in a subsequent paper.
This constraint is a useful addition to the host of other 
constraints emerging from cosmological observations \cite{lowerN}.

We thank Josh Frieman, Will Kinney, Rocky Kolb, and Andrew
Liddle for useful conversations.
We are indebted to Christian Armendariz-Picon for pointing out an 
important loophole in an earlier
version of our paper. This work is supported by the DOE and its OJI program, 
by NASA grant NAG5-10842, and by NSF Grant PHY-0079251.

\newcommand\spr[3]{{\it Physics Reports} {\bf #1}, #2 (#3)}
\newcommand\sapj[3]{ {\it Astrophys. J.} {\bf #1}, #2 (#3) }
\newcommand\sprd[3]{ {\it Phys. Rev. D} {\bf #1}, #2 (#3) }
\newcommand\sprl[3]{ {\it Phys. Rev. Letters} {\bf #1}, #2 (#3) }
\newcommand\np[3]{ {\it Nucl.~Phys. B} {\bf #1}, #2 (#3) }
\newcommand\smnras[3]{{\it Monthly Notices of Royal
	Astronomical Society} {\bf #1}, #2 (#3)}
\newcommand\splb[3]{{\it Physics Letters} {\bf B#1}, #2 (#3)}


\begin{thebibliography}{99}
\bibitem{horizon} Following common practice, we loosely use 
the term horizon interchangeably with the more proper term, the Hubble radius
or apparent horizon.
\bibitem{othermodels} For new inflation (i.e. small field) with 
a potential of the form $V = \Lambda^4 [1 - (\phi/\mu)^p]$,
where $p > 2$, and $\phi \ll \mu$, it can be shown that
\begin{eqnarray}
\epsilon = \left[{2^{2p+2} \pi^p \over p^2 (p-2)^{2p-2}} 
\left( {\mu \over m_{\rm pl.}} \right)^{2p} N^{2-2p} \right]^{1/(p-2)} 
\nonumber 
\end{eqnarray}
and $\eta = - N^{-1} (p-1)/(p-2)$, 
which determine $n_s$, $n_T$, and $A_T/A_S$ through 
equation (\ref{basics}). 
Hybrid inflation, where one field provides the vacuum energy floor
while another rolls, can be treated effectively as a single-field model:
$V = \Lambda^4 [1 + (\phi/\mu)^p]$.
In this case, $\epsilon$ is the same as in new inflation, while
$\eta$ has the opposite sign. See \cite{kinney} for a more thorough classification.
\bibitem{andrew} This constraint has been derived independently by
\cite{ll}.
\bibitem{tensor} Obtaining the constraint on $H_*$ from the literature 
is not entirely straightfoward, partly
because of different definitions of the tensor to scalar ratio $r$.
Here, we obtain our constraints from \cite{wmap} as follows: 
$(16/\pi) (H_*^2/m_{\rm pl.}^2) < 1.28 \times 2.95 \times 10^{-9} 
\times (0.71 + 3 \times 0.1)$, where $0.71 + 3 \times 0.1$ is the 3 $\sigma$ 
upper limit on the scalar amplitude (at 
$k = 0.002$ Mpc$^{-1}$). The factor of $1.28$ gives the upper limit to $r$ as defined by
\cite{wmap}.
\end{thebibliography}
\end{document}